\documentclass[twocolumn]{aastex62}
\usepackage{ulem}
\received{August 12, 2019}
\revised{December 20, 2019}
\accepted{January 7, 2020}
\submitjournal{ApJ}

\shorttitle{Fe-rich Knot in Kepler's SNR}
\shortauthors{Sato et al.}

\begin{document}

\title{A Nucleosynthetic Origin for the Southwestern Fe-rich Structure in Kepler's Supernova Remnant}

\correspondingauthor{Toshiki Sato}
\email{toshiki.sato@riken.jp}

\author[0000-0001-9267-1693]{Toshiki Sato}
\affil{RIKEN, 2-1 Hirosawa, Wako, Saitama 351-0198, Japan}
\affil{NASA Goddard Space Flight Center, 8800 Greenbelt Road, Greenbelt, MD 20771, USA}
\affil{Department of Physics, University of Maryland Baltimore County, 1000 Hilltop Circle, Baltimore, MD 21250, USA}

\author[0000-0003-0894-6450]{Eduardo Bravo}
\affiliation{E.T.S. Arquitectura del Valle\`s, Universitat Polite\`cnica de Catalunya, Carrer Pere Serra 1-15, 08173 Sant Cugat del Valle\`s, Spain}

\author[0000-0003-3494-343X]{Carles Badenes}
\affiliation{Astrophysics and Cosmology Center (PITT PACC), Department of Physics and Astronomy and Pittsburgh Particle Physics, University of Pittsburgh, 3941 O’Hara Street, Pittsburgh, PA 15260, USA}
\affiliation{Institut de Cie\`ncies del Cosmos (ICCUB), Universitat de Barcelona (IEEC-UB), Mart\'i Franques 1, E-08028 Barcelona, Spain}

\author[0000-0002-8816-6800]{John P.Hughes}
\affiliation{Department of Physics and Astronomy, 
Rutgers University, 136 Frelinghuysen Road, Piscataway, 
NJ 08854-8019, USA}

\author[0000-0003-2063-381X]{Brian J. Williams}
\affiliation{NASA, Goddard Space Flight Center, 8800 Greenbelt Road, Greenbelt, MD 20771, USA}

\author[0000-0002-5092-6085]{Hiroya Yamaguchi}
\affiliation{Department of High Energy Astrophysics, Institute of Space and Astronautical Science (ISAS), Japan Aerospace Exploration Agency (JAXA), 3-1-1 Yoshinodai, Sagamihara, 229-8510, Japan}




\begin{abstract}
{\it Chandra} X-ray observations of Kepler's supernova remnant indicate the existence of a high speed Fe-rich ejecta structure in the southwestern region. We report strong K-shell emission from Fe-peak elements (Cr, Mn, Fe, Ni), as well as Ca, in this Fe-rich structure, implying that those elements could be produced in the inner area of the exploding white dwarf. We found Ca/Fe, Cr/Fe, Mn/Fe and Ni/Fe mass ratios of 1.0--4.1\%, 1.0--4.6\%, 1--11\% and 2--30\%, respectively. In order to constrain the burning regime that could produce this structure, we compared these observed mass ratios with those in 18 one-dimensional Type Ia nucleosynthesis models (including both near-$M_{\rm Ch}$ and sub-$M_{\rm Ch}$ explosion models). 
The observed mass ratios agree well with those around the middle layer of incomplete Si-burning in Type Ia nucleosynthesis models with a peak temperature of $\sim$(5.0--5.3)$\times$10$^{9}$ K and a high metallicity, Z $>$ 0.0225. Based on our results, we infer the necessity for some mechanism to produce protruding Fe-rich clumps dominated by incomplete Si-burning products during the explosion. We also discuss the future perspectives of X-ray observations of Fe-rich structures in other Type Ia supernova remnants.


\end{abstract}



\section{Introduction} \label{sec:intro}
The nuclear reaction of carbon ignition at the central core of white dwarfs (WDs) is thought to lead to a thermonuclear runway and a Type Ia supernova (SNe Ia). Even though they are extremely important phenomena in the universe (e.g., standard candles for cosmology, major sources of Fe), many fundamental aspects of these explosions remain obscure (e.g., the explosion mass, the progenitor system). 

There are two major channels that are thought to lead to SN Ia explosions. One is the single-degenerate (SD) scenario, where a WD obtains materials from a non-degenerate companion to increase the mass until it explodes \citep[e.g.,][]{1973ApJ...186.1007W}.
The other is the double-degenerate (DD) scenario, where the explosion comes from a binary of two WDs. \citep[e.g.,][]{1984ApJ...277..355W}. The SD scenarios usually assume that WDs explode when their mass gets close to the Chandrasekhar limit
(M$_{\rm Ch}$ $\approx$ 1.4 $M_\odot$) \citep[e.g.,][]{1999ApJS..125..439I} The classical DD scenario was also considered as the explosion of near-M$_{\rm Ch}$ WDs where the total masses of merging WDs exceed the Chandrasekhar limit. On the other hand, some theoretical studies have indicated that WD mergers are difficult to be SNe Ia, but instead collapse to neutron stars \citep[e.g.,][]{1985A&A...150L..21S}. Thus, recent DD
scenarios involve the explosion of sub-M$_{\rm Ch}$ WDs \citep[e.g.,][]{2012ApJ...747L..10P,2018ApJ...854...52S}.

Kepler's supernova \citep[SN 1604;][]{2017hsn..book..139V} is one of the most well-studied young SNRs in the Galaxy. The general consensus that the remnant is a Type Ia SN is based largely on X-ray observations showing shocked ejecta with strong Fe emission and a near absence of O emission \citep[e.g.,][]{2007ApJ...668L.135R}. The remnant is thought to be
interacting with a dense circumstellar medium \citep[e.g.,][]{2013ApJ...764...63B,2015ApJ...808...49K,blair07,williams12}, 
which supports the SD scenario as the Kelper's origin. On the other hand, an obvious surviving companion star that would represent the strongest evidence of the SD scenario has not yet been found \citep[][]{2014ApJ...782...27K,2017ApJ...845..167S,2018ApJ...862..124R}, therefore the origin is still under debate.

One important observational signature that would offer clues to its origin would be a large amount of Fe-peak elements in the remnant. \cite{2012ApJ...756....6P} performed hydrodynamical and spectral modeling of Kepler's SNR and showed that the X-ray spectrum is consistent with an explosion that produced $\sim$1 $M_{\odot}$ of $^{56}$Ni, ruling out the subenergetic models ($^{56}$Ni mass = 0.3 $M_{\odot}$), suggesting that the remnant was an SN 1991T-like event \citep[see also][]{2015ApJ...808...49K}. 
On the other hand, the light curve recorded $\sim 400$ yrs ago indicates a normal SN Ia at a distance of 5$\pm$0.7 kpc \citep{2017ApJ...842..112R}. Thus, Kepler's supernova seems to have been either a normal or bright SN Ia.

\begin{figure}[t!]
 \begin{center}
  \includegraphics[bb=0 0 740 555, width=9cm]{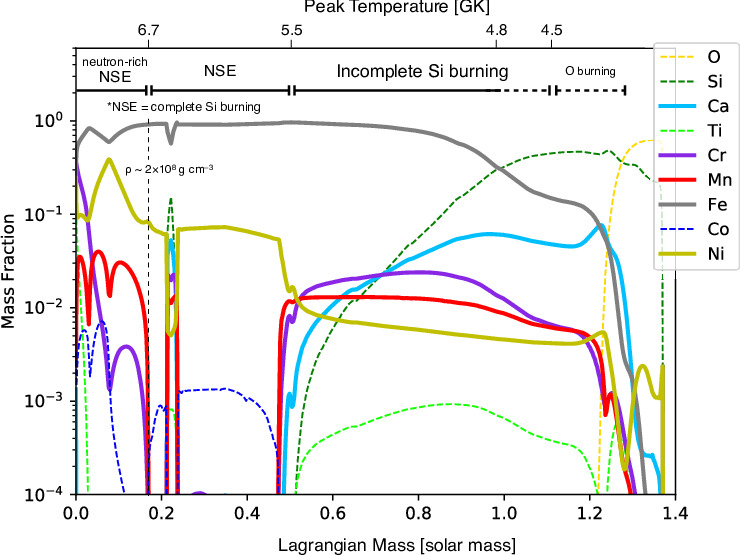}
 \end{center}
\caption{An example of the one-dimensional Type Ia nucleosynthesis models in \cite{2019MNRAS.482.4346B}. The model assumed a delayed detonation with the transition density, $\rho = 2.8 \times 10^{8}$ g cm$^{-3}$ and the metallicity of 1.61 $Z_{\odot}$. The Fe-rich ejecta must be produced at inner area with the peak temperature above $\sim$4.8$\times$10$^{9}$ K. A discontinuity around 0.2 $M_\odot$ in the abundances is a transition point from deflagration to detonation}.
\label{fig:nucleosynthesis}
\end{figure}

To provide new information on the explosion, we focus on a peculiar ``Fe-rich'' knot at the southwestern region for this study \citep[labeled as ``hand" in][]{2004A&A...414..545C}. The Fe-rich knot is only slightly decelerated (expansion index, $m \sim$0.5--0.8) with a very high 3D velocity of $\sim$5,000--8,000 km/s \citep{2017ApJ...845..167S}. The high velocity implies the knot may be located at the SNR front (although it seems to be a little inside in the 2D image due to projection effects), and the Fe-rich composition means the knot may have been synthesized at the deep side of the SN Ia. This is because the Fe-rich ejecta can be achieved only at inner area of SNe Ia with the peak temperature above $\sim$4.8$\times$10$^{9}$ K (see Figure \ref{fig:nucleosynthesis}). The existence of clumpy Fe-rich structures has been proposed also in some other Type Ia SNe and SNRs \citep[Tycho's SNR, SN 1885, SNR 0509-67.5;][]{1995ApJ...441..680V,2015ApJ...804..140F,2019MNRAS.483.1114B,2019PhRvL.123d1101S}. We infer those Fe-rich knots were pushed out from the inner layer to the SN surface by some sort of asymmetric effect during the explosion \citep[e.g. instabilities, bouyancy;][]{2000astro.ph..8463K,2009Natur.460..869K}, while preserving the information on the nucleosynthesis and explosion (see a detail discussion in section 4.1).

There are three different burning regimes where Fe is produced (Figure \ref{fig:nucleosynthesis}): incomplete Si burning, nuclear statistical equilibrium (NSE), and neutron-rich NSE (n-NSE). 
In these regimes, the synthesis of the Fe-peak elements is characterized by three types of neutronization \citep[see a summary in][]{2017ApJ...843...35M}. The neutronization has information on the progenitor WDs (e.g., metallicity, progenitor mass). Thus, evaluating the neutronization in the Fe-rich ejecta provides unique information about the progenitor \citep[e.g.,][]{2008ApJ...680L..33B,2013ApJ...767L..10P,2015ApJ...801L..31Y}, after accounting for radioactive decays that transmute the freshly syntesized nuclei into their stable isobars.

In the incomplete Si burning and NSE regimes, the yield of neutron-rich species (e.g., those that end as $^{55}$Mn, $^{58}$Ni, $^{59}$Co, after beta-decays) is mainly controlled by the pre-explosion neutron excess carried by $^{22}$Ne in the WD, which in turn is set mainly by the metallicity of the WD progenitor \citep[e.g.,][]{2003ApJ...590L..83T,2008ApJ...680L..33B}. Only in the case of near-M$_{\rm Ch}$ WDs is the innermost region at $\lesssim$0.2 M$_\odot$ (i.e., at $\rho > 2 \times 10^8$ cm$^{-3}$ g) consumed in the n-NSE regime \citep[e.g.,][]{1986A&A...158...17T,1999ApJS..125..439I}, where density-driven electron capture generates a neutron excess independent of the progenitor metallicity. In addition, the composition of the NSE layers may be altered during the explosion in case of a high-entropy freeze-out (at $\rho < 10^8$ g cm$^{-3}$) because of the presence of light particles, which is called $\alpha$-rich freeze-out, as opposed to the normal freeze-out. Finally, ``carbon simmering'' is also an important neutronization process in SN Ia progenitors \citep[e.g.,][]{2008ApJ...673.1009P}. In near-$M_{\rm Ch}$ progenitors, carbon can ignite around the convective core without a thermonuclear runaway. This convective carbon burning core is active for $\sim$10$^3$ yr prior to the explosion. During this simmering, electron captures occur on the products of carbon burning, which decreases the electron fraction. Here, both carbon simmering and n-NSE require near-$M_{\rm Ch}$ WDs, therefore a hard observational limit on the neutron excess in the Fe-rich ejecta will help us to discriminate between near-$M_{\rm Ch}$ and sub-$M_{\rm Ch}$ SNe Ia.


\cite{2017ApJ...834..124Y} demonstrated constraints on the burning regime for the Fe-rich structure using a similar Fe-rich knot seen in Tycho's SNR. X-ray imaging with the {\it Einstein} and {\it ROSAT} missions have already reported the existence of Si-rich and Fe-rich ejecta knots at the southeastern region of the remnant \citep{1995ApJ...441..680V}. Interestingly, the {\it Suzaku} spectrum of the Fe-rich knot show no emission from the other Fe-peak elements (Cr, Mn, Ni). Using the mass fraction among the Fe-peak elements, they concluded that either incomplete Si burning or an $\alpha$-rich freeze-out regime would be its origin.

As done by \cite{2017ApJ...834..124Y}, we can specify the burning regimes for the local Fe-rich structures in Type Ia SNRs using knowledge of the nucleosynthesis and the observed abundances. In this paper, we investigate the elemental abundance in this Fe-rich structure of Kepler's SNR for the first time.


\begin{figure}[t!]
 \begin{center}
  \includegraphics[bb=0 0 507 462, width=8cm]{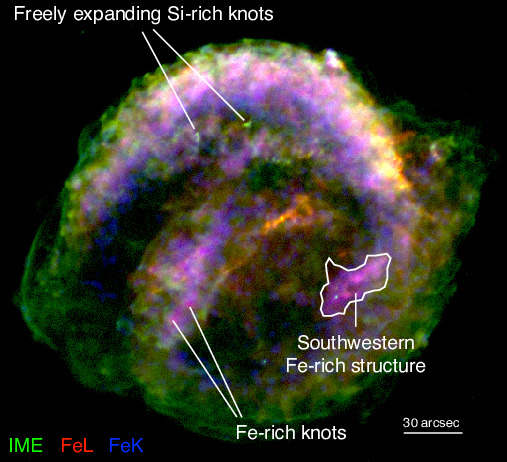}
 \end{center}
\caption{{\it Chandra} X-ray image of Kepler's SNR, combining images in green, red and blue made from energy bands of 1.76--4.2 keV (emission from IMEs including Si, S, Ar, Ca), 0.72--1.3 keV (Fe L-shell emission) and 6.2--6.7 keV (Fe K$\alpha$), respectively. The scale bar indicates a size of 30 arcsec. The white contour region where the Fe emissions are dominant was chosen using the ratio image between the Fe L-shell and IME emissions.}
\label{fig:f1}
\end{figure}

\section{Observations and Analysis} \label{sec:obs}
The {\it Chandra} ACIS-S instrument performed a deep observation of Kepler's SNR in 2006 (PI: S. Reynolds). The net exposure time is 741.0 ks. We reprocessed all the level-1 event data, applying the standard data reduction with CALDB version 4.7.8. In this process, we use a custom pipeline based on ``\verb"chandra_repro"'' in CIAO version 4.10.

Figure \ref{fig:f1} shows a three-color image of Kepler's SNR where the Fe-rich regions are emphasized by red (Fe L-shell emission) and blue (Fe K$\alpha$) colors. The Fe-rich structures seem to be mainly concentrated in the central, northern and southwestern regions. In case of the central region, small Fe-rich knots are adjacent to Si-rich knots (seen in green color). The northern bright Fe-rich shell may come from overdensities in the ambient medium \citep{2004A&A...414..545C}. The Fe-rich structure the southwest is composed of some knotty structures, which are more widely spread than those in the central region (the area is $\sim$3\% of the remnant). In this study, we focus only on the bright southwest Fe-rich region.

\begin{figure}[t!]
 \begin{center}
  \includegraphics[bb=0 0 493 490, width=8cm]{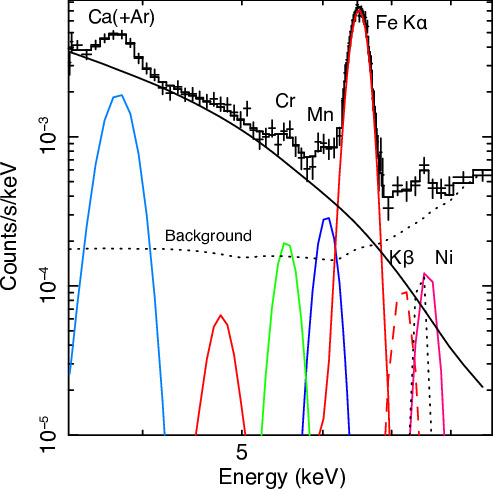}
 \end{center}
\caption{X-ray spectrum of the Kepler Fe-rich knot over the 3.4--9 keV band fitted with a phenomenological model including Gaussian lines for Ca (light blue), Cr (green), Mn (blue), Fe (red) and Ni (magenta). The power-law model (black curve) well describes the continuum emission (which includes both non-thermal synchrotron and thermal bremsstrahlung radiation). The modeled background spectrum is plotted with the black dotted line.}
\label{fig:f2}
\end{figure}

\begin{figure}[t!]
 \begin{center}
  \includegraphics[bb=0 0 637 633, width=8cm]{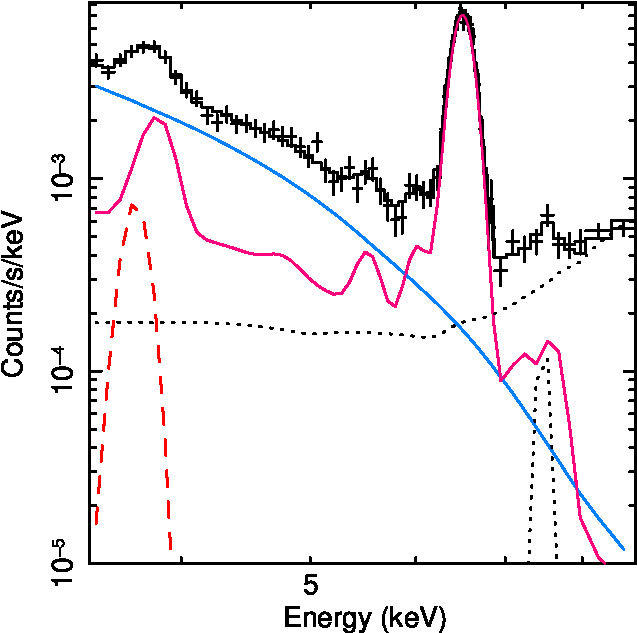}
 \end{center}
\caption{The same as Figure \ref{fig:f2}, but now using a thermal plasma model to fit the line emission. The colored lines show the components of the best-fit model: the {\tt vvpshock} thermal plasma model (magenta), a power law (blue) and a Gaussian line for Ar K$\beta$ (dashed red line).}
\label{fig:pshock}
\end{figure}

We extracted the spectrum (Figure \ref{fig:f2}) from the white contour region in Figure \ref{fig:f1}.
Here, we use the X-ray spectrum above 3.4 keV where we can see the emission from the Fe-group elements and Ca. We modeled the background spectrum using the surrounding blank sky (dotted lines in Figure~\ref{fig:f2}). We checked the validity of this model using 5 different background regions surrounding the remnant. As a result, we found our background model very well explains all the background spectra ($\chi^2$/d.o.f = 0.94--1.07 in 5--9 keV band). A model consisting of a power law $+$ several Gaussian models  $+$ background model fits the spectrum well ($\chi^2$/d.o.f = 42.16/55), and we found the spectrum has strong K-shell emission from the Fe-peak elements (Cr, Mn, Fe, Ni) where the significance for each element is above 2$\sigma$ (Table \ref{tab:Gafit}). The centroid energies of all the lines are slightly higher ($\sim$30--70 eV) than those in the remnant as a whole \citep{2013ApJ...767L..10P}, consistent with a spectrum blueshifted by several thousand km s$^{-1}$ as reported in \cite{2017ApJ...845..167S}. As we show below, spectral fits using a physical NEI model also require a comparable blueshift for the Fe-rich knot.

\begin{table}[h]
\scriptsize
\caption{Best-fit Parameters of Gaussian Models in the SW knot$^\star$.}
\begin{center}
\begin{tabular}{cccccc}
\hline
line ids        &     Centroids [keV]       & Flux                      & significance$^\dagger$      \\
                &      [keV]                & [10$^{-7}$ ph/cm$^2$/s]   &                   \\ \hline
Ca(+Ar)         &   3.79$\pm$0.03           & 16.0$^{+5.0}_{-3.8}$                 & 6.6$\sigma$       \\
Cr K$\alpha$    &   5.53$^{+0.09}_{-0.10}$  & 2.5$\pm$1.9               & 2.1$\sigma$       \\
Mn K$\alpha$    &   6.06$^{+0.07}_{-0.06}$  & 4.9$\pm$2.2               & 3.6$\sigma$       \\
Fe K$\alpha$    &   6.515$\pm$0.006         & 153$\pm$7                 & 35.7$\sigma$       \\
Fe K$\beta$     &   7.19$^{+0.19}_{-0.15}$  & 4.2$\pm$3.7               & 1.8$\sigma$       \\
Ni K$\alpha$    &   7.60$^{+0.32}_{-0.14}$  & 8.2$^{+6.4}_{-6.3}$       & 2.1$\sigma$       \\ \hline
\end{tabular}
\begin{flushleft}
\item $^\star$ the error shows 90\% confidence level ($\Delta \chi^2$ = 2.7). The line widths except for the Ca(+Ar) line are linked to that of Fe K$\alpha$, $\sigma=$ 103$\pm$6 eV. The line width of the Ca(+Ar) line is $\sigma=$ 112 (+37/--31) eV.
\item $^\dagger$ the detection significance is estimated at the best-fit centroid energy.
\end{flushleft}
\label{tab:Gafit}
\end{center}
\end{table}

\begin{table*}[t]
\caption{Best-fit Parameters of pshock Models in the SW knot$^\star$.}
\begin{center}
\begin{tabular}{cccccccc}
\hline
\multicolumn{6}{l}{Fitting parameters of pshock model}\\
$kT_{\rm e}$    &     [Ca/Fe]/[Ca/Fe]$_{\odot}$     & [Cr/Fe]/[Cr/Fe]$_{\odot}$     &     [Mn/Fe]/[Mn/Fe]$_{\odot}$     &    [Ni/Fe]/[Ni/Fe]$_{\odot}$      & $nt$                      & z             \\ 
~[keV]          &                                   &                               &                                   &                                   &  [10$^{10}$ cm$^{-3}$ s]  &  [10$^{-2}$]  \\
4               &    0.6$^{+0.4}_{-0.3}$            & 2.8$^{+1.6}_{-1.7}$           & 3.9$^{+16.7}_{-1.8}$              & 4.1$^{+3.5}_{-3.4}$               & 1.8$\pm$0.3               & $-1.12 \pm 0.27$  \\
6               &    0.7$^{+0.4}_{-0.3}$            & 3.0$^{+1.8}_{-1.9}$           & 3.6$^{+9.8}_{-1.4}$               & 3.7$\pm$3.0                       & 1.7$^{+0.3}_{-0.2}$       & $-1.12^{+0.21}_{-0.22}$  \\
8               &    0.8$\pm$0.4                    & 3.1$^{+1.9}_{-2.0}$           & 3.7$^{+14.7}_{-1.5}$              & 3.3$^{+2.5}_{-2.9}$               & 1.8$\pm$0.3               & $-1.12^{+0.55}_{-0.02}$  \\ \hline
\multicolumn{6}{l}{Estimated mass ratios}\\
 $kT_{\rm e}$   &     Ca/Fe                         &    Cr/Fe                      &      Mn/Fe                        &           Ni/Fe                   & & & \\
  ~[keV]        &      [\%]                         &       [\%]                    &      [\%]                         &           [\%]                    & &  &\\
4               &   2.2$\pm 1.2$                    & 2.6$\pm 1.5$                  & 2.0$^{+8.6}_{-0.9}$               & 16$^{+14}_{-13}$                  &&&\\
6               &   2.5$^{+1.3}_{-1.2}$             & 2.7$^{+1.7}_{-1.8}$           & 1.8$^{+5.0}_{-0.7}$               & 15$\pm$12                         &&&\\
8               &   2.7$^{+1.4}_{-1.3}$             & 2.9$\pm 1.8$                  & 1.9$^{+7.6}_{-0.8}$               & 13$^{+10}_{-12}$                  &&&\\ \hline
\end{tabular}
\item $^\star$ The error shows 90\% confidence level ($\Delta \chi^2$ = 2.7). The lower ionization state for the pshock model is fixed to 1$\times$10$^{9}$ cm$^{-3}$ s. We used the solar abundances of \cite{1989GeCoA..53..197A}. The broadening width is $\sigma_{\rm gsmooth} = 90_{-6}^{+12}$ eV at 6 keV.
\label{tab:pshockfit}
\end{center}
\end{table*}

Figure \ref{fig:pshock} and Table \ref{tab:pshockfit} show the best-fit results from a thermal plasma model (vvpshock model) plus a power-law model (i.e., a model for non-thermal emission). To fit lines broadened by thermal broadening and/or multiple velocity components, we also used a gsmooth model in Xspec. We assumed that the plasma parameters (e.g., ionization state, temperature, redshift) for each element are identical. We added an additional Gaussian line at $\sim$3.7 keV (broken red line in the figure) to account for Ar K$\beta$. It is difficult to constrain the electron temperature from an X-ray spectrum containing nonthermal emission, so we derive all plasma parameters using different three different electron temperatures: 4, 6, and 8 keV. Fits to the Fe-K$\alpha$ line seen by {\it Suzaku} showed an ionization timescale of $n_{\rm e}t \approx 2 \times 10^{10}$  cm$^{-3}$ s and an electron temperature in the range $kT_{\rm e} =$ 3--8 keV for the Fe ejecta \citep{2013ApJ...767L..10P}, which are roughly consistent with those in our plasma models (see Table \ref{tab:pshockfit}). We found all the line structures in the Gaussian model are well explained by the thermal plasma models, where the goodness of fits are $\chi^2$/d.o.f = 40.33/56 for 4 keV, $\chi^2$/d.o.f = 40.27/56 for 6 keV and $\chi^2$/d.o.f = 40.49/56 for 8 keV. The plasma parameters are only minimally changed by the choice of different electron temperatures. This is because the line emissivities for each element do not change significantly in these temperature ranges. In the case of the abundances, the differences on the best-fit values are $\sim$3--14\%. The spectrum shows a blue-shift velocity of $\sim$3,400 km s$^{-1}$, which is a little smaller than those of SW1 and SW2 in \cite{2017ApJ...845..167S}. The likely reason for the smaller velocity is the integration of larger area in this study and/or the difference of the fitting energy range between them. 

We summarize the estimated mass ratios from the X-ray spectrum in the Fe-rich structure of Kepler's SNR in Table \ref{tab:pshockfit}. The best-fit mass ratios, Ca/Fe, Cr/Fe and Mn/Fe show $\sim$2--3\%, which is roughly consistent with the mass fraction around the incomplete Si burning regime (at the Lagrangian mass $\sim$ 0.7$M_{\odot}$ in Figure \ref{fig:nucleosynthesis}). On the other hand, the best-fit mass ratio of Ni/Fe is above 10\%, although the error is also large (the lower limit is $\sim$1--3\%) due to the low photon statistics and the high background levels. In order to produce Ni/Fe $>$ 10\%, the NSE and n-NSE would be reasonable regimes (or incomplete Si burning with an unrealistically large metallicity). More accurate future measurements of the Ni line will be useful to understand the burning regimes and the progenitor metallicity. Detailed discussions on the burning regimes are given in following sections. Here, we assumed atomic mass ratios of 0.714, 0.929, and 0.982 for Ca/Fe, Cr/Fe and Mn/Fe, respectively, where we calculated the mass number ratio among $^{40}$Ca, $^{52}$Cr, $^{55}$Mn and $^{56}$Fe.

\section{Comparison with Nucleosynthesis Models} \label{sec:nucleosynthesis} 
We found strong K-shell emission from the Fe-peak elements in the southwestern Fe-rich knots of Kepler's SNR. The mass fractions we measured in the previous section contain information on the nucleosynthesis that occurred during the explosion that led to the remnant. 
We are concerned with elemental abundances resulting from the supernova explosion after radioactive decays. Although the abundances of most iron group elements are dominated by a few isotopes, we need to consider all possible origins for each element, since we measure them in a small region of the ejecta. For instance, consider the case of manganese, which in the remnant is $^{55}$Mn. In Figure \ref{fig:nucleosynthesis}, there can be seen three peaks in the Mn abundance close to the WD center: the first one comes from the explosive synthesis of $^{55}$Mn directly, the second one from that of $^{55}$Fe, and the third one from $^{55}$Co, and each one of these isobars reflects a different neutron excess. To address these complications we rely on supernova explosion models.

In this section, we investigate the origin of the Fe-rich structure by comparing to mass fractions determined from one-dimensional nucleosynthesis models of thermonuclear SNe.
Although the Fe-rich structure likely has some contamination from fainter overlying outer layers due to geometric projection, we assume the spectrum and therefore the mass fractions we derive are dominated by the bright Fe-rich structure (see Appendix). In addition, we also assume that the mass fractions in the Fe-rich structure can be traced to a  specific peak temperature within a narrow range of the burning layers. Here we use the models in \cite{2019MNRAS.482.4346B} \citep[see also][]{2018ApJ...865..151M} which do not include effects of carbon simmering or radial mixing of ejecta layers. The Type Ia models we use are summarized in Table \ref{tab:list}. Kepler's supernova is thought to have been a normal or bright SN Ia \citep{2012ApJ...756....6P,2015ApJ...808...49K,2017ApJ...842..112R}, thus we choose models whose $^{56}$Ni yields lead to normal or bright SN Ia with different metallicities: Z = 0.009, 0.0225, 0.0675 (0.67 Z$_{\odot}$, 1.68 Z$_{\odot}$, 5.04 Z$_{\odot}$) \citep{2009ARA&A..47..481A}.

We use the four observed mass ratios in Table~\ref{tab:pshockfit} for Ca/Fe, Cr/Fe, Mn/Fe, and Ni/Fe to constrain the burning regime for the Fe-rich structure. Ca is produced nearly entirely within the incomplete Si burning region; Ni is also produced in this layer, but is produced even more abundantly in the deeper NSE (complete Si burning) and n-NSE layers (see Fig.~\ref{fig:nucleosynthesis}). Cr and Mn are present in both the incomplete Si burning and n-NSE layers. 

\begin{figure}[h!]
 \begin{center}
  \includegraphics[bb=0 0 975 776, width=8cm]{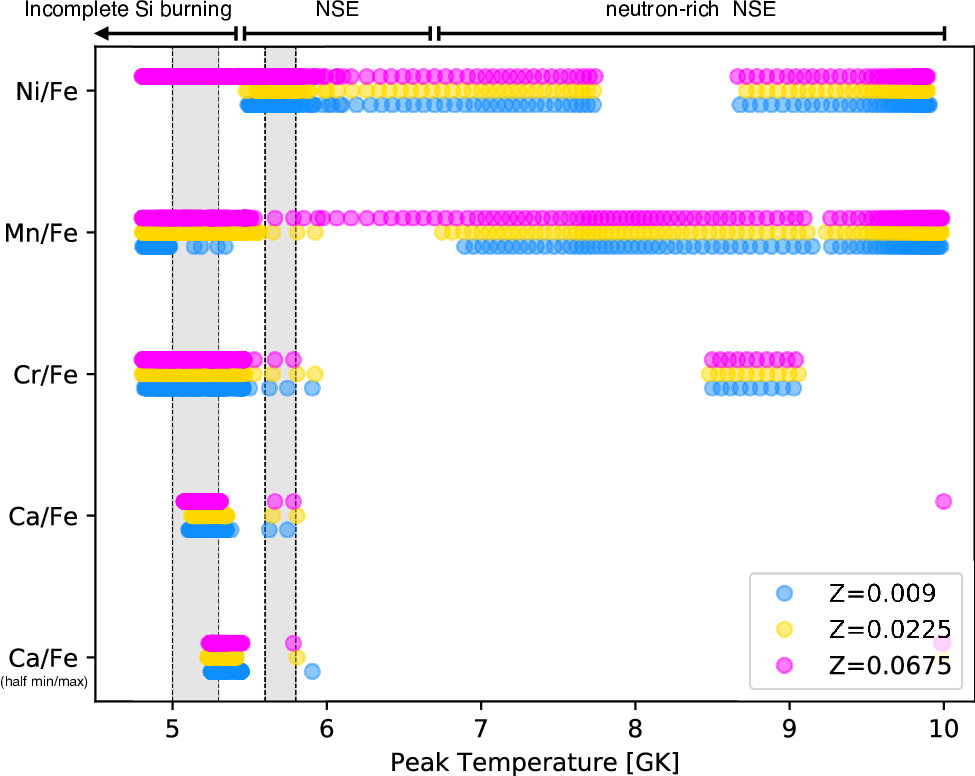}
 \end{center}
\caption{Peak temperatures from the Type Ia nucleosynthesis models (along the horizontal axis) consistent with each of our four measured mass ratios (along the vertical axis). This is for the delayed detonation model with transition density, $\rho = 2.8 \times 10^{7}$ g cm$^{-3}$ and progenitor metallicities of Z = 0.009 (blue), Z = 0.0225 (yellow) and Z = 0.0675 (magenta).  For each mass ratio, peak temperature and metallicity, we plot a point if the specific mass ratio from the model is consistent (at 90\% confidence) with the measured value.  The gray shaded vertical bands mark the regimes allowed by all four mass ratio measurements. See text for more details. At the last line, we artificially reduce the Ca abundance by half.
}
\label{fig:restricted_T}
\end{figure}

\begin{table}[h]
\scriptsize
\caption{Type Ia nucleosynthesis models and the constrained peak temperatures.}
\begin{center}
\begin{tabular}{cccc}
\hline
\multicolumn{4}{l}{\bf near-$M_{\rm Ch}$ delayed-detonation models} \\
$\rho_{\rm DDT}$    &   Z       &   $M$($^{56}$Ni)  &  constrained $T_{\rm peak}$\\
~[g cm$^{-3}$]      &           &  [$M_{\odot}$]    &      [GK]                 \\ \hline
2.4$\times$10$^7$   &   0.009   &   0.704           &     ---                   \\
2.4$\times$10$^7$   &   0.0225  &   0.663           &     ---                   \\
2.4$\times$10$^7$   &   0.0675  &   0.549           &  5.10--5.34, 5.56--5.72   \\
2.8$\times$10$^7$   &   0.009   &   0.765           &     ---                   \\
2.8$\times$10$^7$   &   0.0225  &   0.721           &     ---                   \\
2.8$\times$10$^7$   &   0.0675  &   0.595           &  5.07--5.31, 5.66--5.78   \\
4.0$\times$10$^7$   &   0.009   &   0.872           &     ---                   \\
4.0$\times$10$^7$   &   0.0225  &   0.824           &     ---                   \\
4.0$\times$10$^7$   &   0.0675  &   0.689           &  5.04--5.27               \\ \hline
\multicolumn{4}{l}{\bf sub-$M_{\rm Ch}$ detonation models} \\
$M_{\rm WD}$        &   Z       &   $M$($^{56}$Ni)  &  constrained $T_{\rm peak}$\\
~[$M_{\odot}$]      &           &  [$M_{\odot}$]    &      [GK]                 \\ \hline
1.06                &   0.009   &   0.680           &     ---                   \\
1.06                &   0.0225  &   0.650           &     ---                   \\
1.06                &   0.0675  &   0.569           &  5.01--5.22               \\
1.10                &   0.009   &   0.781           &     ---                   \\
1.10                &   0.0225  &   0.748           &     ---                   \\
1.10                &   0.0675  &   0.653           &  4.99--5.20               \\
1.15                &   0.009   &   0.901           &     ---                   \\
1.15                &   0.0225  &   0.865           &     ---                   \\
1.15                &   0.0675  &   0.753           &  4.97--5.18               \\
\hline
\end{tabular}
\label{tab:list}
\end{center}
\end{table}

\begin{figure*}[h!]
 \begin{center}
  \includegraphics[width=16cm, bb=0 0 1020 1173]{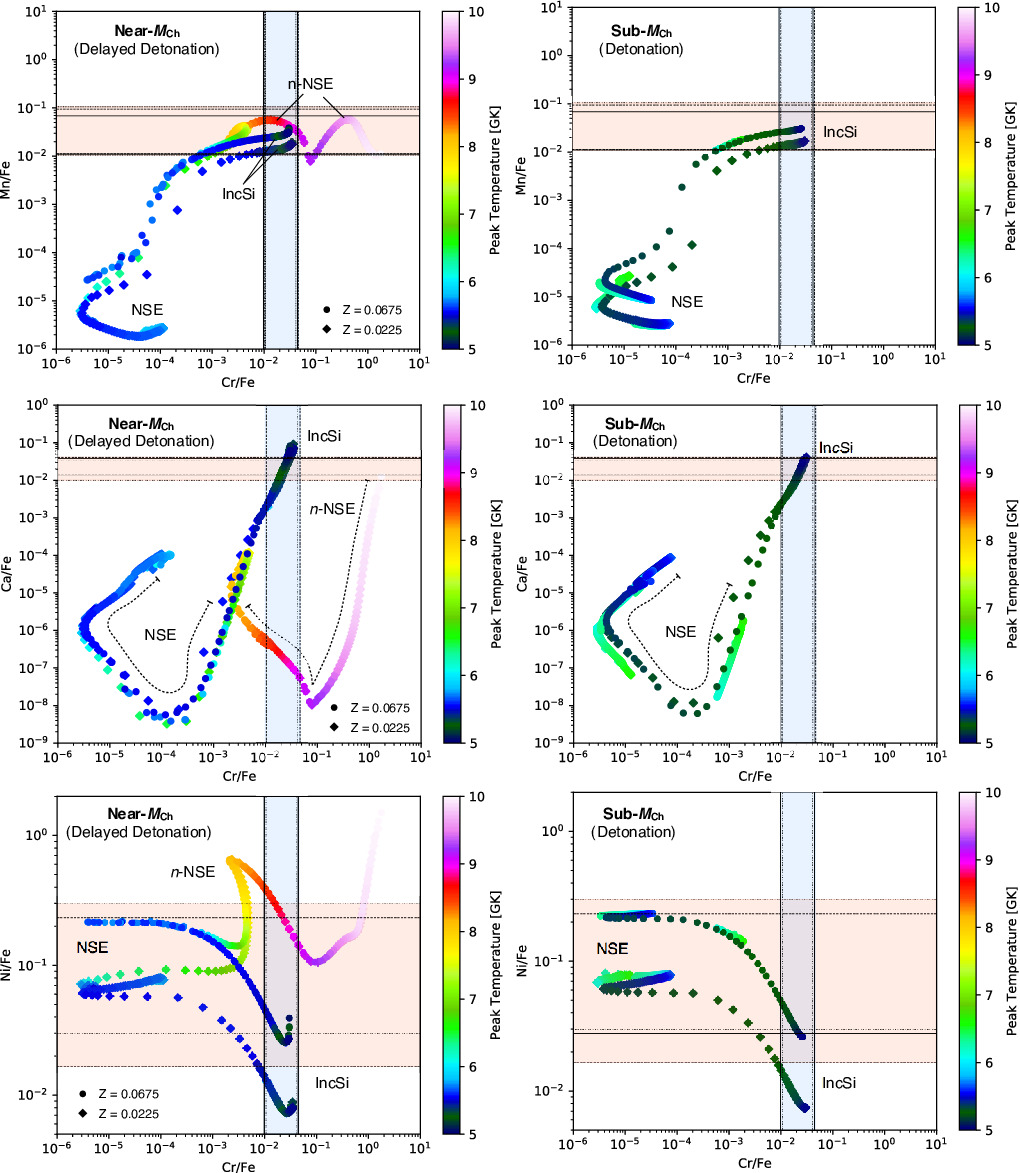}
 \end{center}
\caption{Comparison between the mass ratio in Type Ia nucleosynthesis models (colored points) and the measured values (vertical and horizontal bands), plotted at the 90\% confidence level.  All mass ratios are with respect to Fe and all panels are plotted with the Cr/Fe mass ratio along the horizontal axis. The top panels show the Mn/Fe mass ratio on the vertical axis, middle panels show Ca/Fe, and the bottom ones show Ni/Fe. 
The panels on the left use the delayed detonation model with transition density, $\rho = 2.8 \times 10^{7}$ g cm$^{-3}$ and progenitor metallicities of Z = 0.0225 (filled diamond symbols) and Z = 0.0675 (filled circle symbols). The panels on the right use the sub-$M_{\rm Ch}$ detonation models with WD mass, $M_{\rm WD} = 1.15 M_{\odot}$ and the same progenitor metallicity values. Note that the peak temperatures in the sub-$M_{\rm Ch}$ detonation models do not extend beyond $\sim$7 GK. The dotted, solid and dashed lines associated with the measurement bands show the 90\% upper/lower limits for the different electron temperatures used in the spectral fits, $kT_{\rm e} =$ 4 keV, 6 keV and 8 keV, respectively.}
\label{fig:SNeIamodel}
\end{figure*}

As a result of the comparison with the Type Ia nucleosynthesis models, we conclude that the Fe-rich structure in Kepler's SNR was produced in the incomplete Si burning regime. 
In Figure \ref{fig:restricted_T} we graphically depict this result in the form of the allowed ranges of peak temperatures that match our four measured mass ratios for the Fe-rich knot using the nucleosynthesis calculations for the delayed detonation models. The models span a temperature range from $4.8\times 10^9$ K to $10^{10}$ K for three different progenitor metallicities (denoted by three different colors: magenta, yellow and blue). For each mass ratio we plot a point if the model ratio matches the measurement (at 90\% confidence). The Ni/Fe mass ratio is the least well constrained by the observations, therefore it shows a large set of allowed temperatures.  Although the Mn/Fe ratio is fairly well constrained, it is produced at the measured ratio over a broad range of temperatures.  Thus these two mass ratios do not tightly constrain the allowed peak temperature or the metallicity (see also Figure \ref{fig:SNeIamodel}, discussed below).  However, the inclusion of the Cr/Fe ratio limits the allowed models to two temperature ranges near $5\times 10^9$ K and $9\times 10^9$ K, and the additional inclusion of the Ca/Fe ratio narrows the acceptable range to the lower temperature part only.  In summary, we find that the ranges of peak temperatures allowed by all four mass ratios lie in only two intervals: $(5.0-5.3)\times 10^9$ K and $(5.6-5.8)\times 10^9$ K (vertical gray bands in Figure \ref{fig:restricted_T}). Both of these peak temperature ranges correspond to incomplete Si burning, but the higher one appears around the deflagration-to-detonation transition (near Lagrangian mass $\sim$ 0.2 $M_{\odot}$ in Figure \ref{fig:nucleosynthesis}). It is difficult to discriminate between these two regimes using only the mass ratios. Yet incomplete Si burning in the narrow transition zone at the higher peak temperatures of $(5.6-5.8)\times 10^9$ K seems to be an unique feature of one-dimensional models. We do not see this feature clearly in the multi-dimensional delayed-detonation models \citep[e.g.,][]{2010ApJ...712..624M,2013MNRAS.429.1156S}. Therefore, we conclude that the lower peak temperature regime may be the more plausible site where the Fe-rich structure originated.

In section \ref{sec:obs}, we could not separate the calcium and argon lines completely from the line structure at 3.8 keV (see Figure \ref{fig:f2} and \ref{fig:pshock}), which may have influence on our interpretation. However, the Ar line is the K${\beta}$ emission, which is much weaker than K${\alpha}$ emissions from Ca. If we remove the Ar line from the model, the abundance of Ca changed by only $\sim$30\%. In addition, even if we reduce the Ca abundance by half, our conclusion would not change (the last line in Figure \ref{fig:restricted_T}). Thus, the uncertainty of the Ar contribution to the 3.8-keV structure is not significant to our results.



\begin{figure}[h!]
 \begin{center}
  \includegraphics[bb=0 0 637 633, width=6cm]{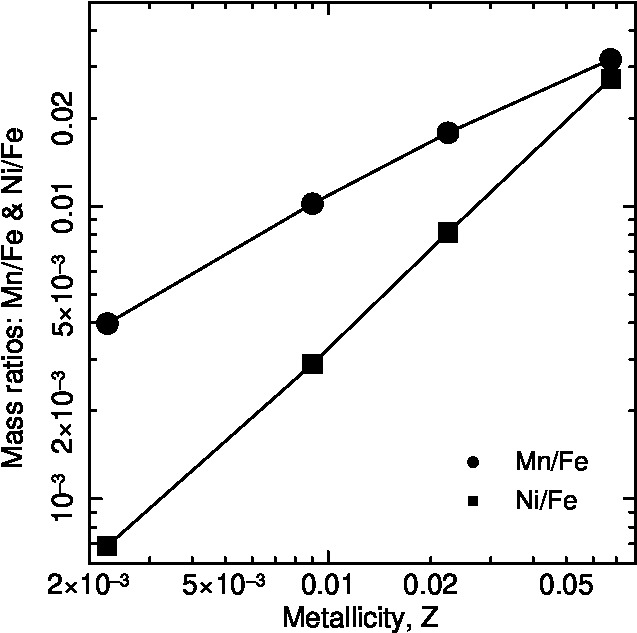}
 \end{center}
\caption{Metallicity dependence of the Mn/Fe (filled circle) and Ni/Fe (filled square) mass ratios in the incomplete Si burning layer (at a peak temperature of 5$\times$10$^{9}$ K) for the delayed detonation model with transition density, $\rho = 2.8 \times 10^{7}$ g cm$^{-3}$.}
\label{fig:MnFe_NiFe}
\end{figure}

We note that our results exclude the models with Z $\leq$ 1.68 Z$_\odot$ (namely the blue and yellow points in Fig.~\ref{fig:restricted_T}), which is generally in agreement with the super-solar metallicity of $\sim$3 Z$_\odot$ inferred from the {\it Suzaku} X-ray spectrum \citep{2013ApJ...767L..10P}. In our case, the lower limit of the Ni/Fe mass ratio ($\gtrsim$ 1\%) requires such a high metallicity.  This can be best seen by examining the Ni/Fe mass ratio curves (bottom panels in Fig.~\ref{fig:SNeIamodel}), which show the most separation between the two metallicity cases especially in the incomplete Si burning zone. There is some separation between the curves for Mn/Fe (top panels) although the $^{58}$Ni production in the incomplete Si burning layer is more sensitive to the metallicity than is the $^{55}$Mn production (see Figure \ref{fig:MnFe_NiFe}). Thus, in SN Ia remnants where we can derive mass ratios for material from the incomplete Si burning layer, the Ni/Fe mass ratio would be very useful for estimating the progenitor's metallicity.

\cite{2019AJ....157...50D} investigated the abundances of interstellar medium clouds in the immediate vicinity of Kepler's SNR, which shows higher element abundances. For example, the carbon and oxygen abundances relative to hydrogen are $\sim$3--5 times higher than those in the sun \citep{2009ARA&A..47..481A}, which would support our result and the {\it Suzaku} observation \citep{2013ApJ...767L..10P}. The authors argued that the abundances in these clouds are representative of the pristine ISM around Kepler's SNR. On the other hand, if the progenitor of this remnant was a fast-moving star \citep[e.g.,][]{1987ApJ...319..885B,2012A&A...537A.139C} with a velocity of $\sim$300 km/s, it may have been likely born in a very different region of the galaxy.

\begin{figure*}[t!]
 \begin{center}
  \includegraphics[bb=0 0 1065 481, width=16cm]{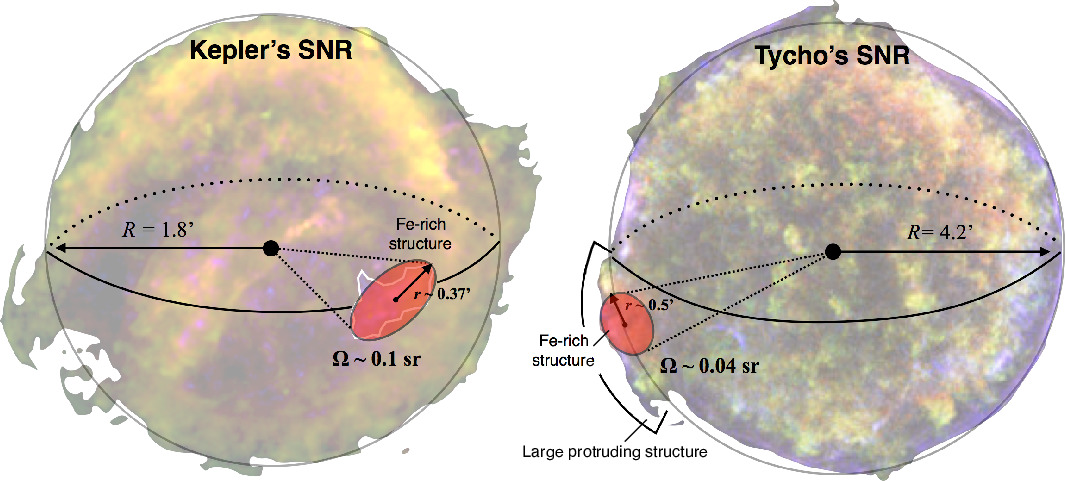}
 \end{center}
\caption{Schematic comparison of the Fe-rich structures in Kepler's and Tycho's SNRs. The solid angles, $\Omega$, were roughly estimated by taking the ratio between the area of a circle with radius $r$ and the square of the remnant's radius, $R$. For the radius, $r$, we take the semi-major axis of the Fe-rich structures (i.e., filled ellipse on each image).}
\label{fig:KeplerTycho}
\end{figure*}

The detailed comparisons of the observed mass ratios (Mn/Fe, Cr/Fe, Ca/Fe and Ni/Fe) with those in both near-$M_{\rm Ch}$ delayed-detonation and sub-$M_{\rm Ch}$ detonation models are shown in Figure \ref{fig:SNeIamodel}. In the case of the sub-$M_{\rm Ch}$ detonation models, the central density of the WD does not exceed $\sim$2$\times$10$^{8}$ g cm$^{-3}$, so electron capture does not occur at the core. This limits the peak temperatures in this model to values below $\sim$7$\times$10$^{9}$ K (see the right panels of Fig.~\ref{fig:SNeIamodel}). Finding evidence in Type Ia SNRs for some Fe-rich structures made at higher peak temperatures would be strong evidence for a near-$M_{\rm Ch}$ explosion, like in the case of 3C 397 \citep{2015ApJ...801L..31Y}. In such a case, the mass ratios of Ca/Fe and Ni/Fe with respect to Cr/Fe would be most useful (i.e., middle and bottom panels of Fig.~\ref{fig:SNeIamodel}). This is because the production of Ca, Cr, and Ni are different from each other and from regime to regime. On the other hand, the Mn/Fe and Cr/Fe mass ratios follow a similar trend (top panel of Fig.~\ref{fig:SNeIamodel}). Mass ratios change in some models by orders of magnitude, so even tight upper limits on faint lines would be helpful to constrain the allowed burning regimes. Future X-ray calorimeter missions such as {\it XRISM} and {\it Athena} will be extremely valuable for this type of study.

\section{Discussion} \label{sec:future}
In the previous sections of this article, we report strong K-shell line emission from the Fe-peak elements Cr, Mn, Fe, and Ni, in addition to Ca in the Fe-rich structure of Kepler's SNR. From the spectral analysis we measure mass ratios with respect to Fe. In section \ref{sec:nucleosynthesis}, we determine the specific burning regime for the Fe-rich structure using the four observed mass ratios and Type Ia nucleosynthesis models. Remarkably, all the observed mass ratios are fully consistent with those at the incomplete Si burning region with peak temperatures of $\sim$(5.0 -- 5.3 ) $\times$ 10$^{9}$ K. Although this implies the action of some mechanisms during the Type Ia explosion to produce distinct Fe-rich clumps from the incomplete-Si-burning regime, we are unable to point to any specific models to explain how such features can form.

Here we summarize the  current understanding from both observational and theoretical studies of mechanisms to produce the Fe-rich structures in SNe Ia (section \ref{sec:origin}). In addition, we expect that the future application of similar approaches using Fe-rich structures will be useful for elucidating more about nucleosynthesis
and clump formation in other Type Ia SNe and SNRs. Therefore we use our Type Ia nucleosynthesis models to discuss the future perspective of X-ray imaging spectroscopy for Type Ia SNRs (section \ref{otherSNR}).

\subsection{How to create Fe-rich clumpy ejecta} \label{sec:origin}

Currently, we are unaware of an obvious mechanism that could produce Fe-rich ejecta knots close to the SN surface. Such knots, in addition to the one studied here in Kepler's SNR, are, however, seen to exist in other Type Ia SNe and SNRs \citep[e.g.,][]{1995ApJ...441..680V,2015ApJ...804..140F,2019MNRAS.483.1114B,2019PhRvL.123d1101S}. In the case of Tycho's SNR, an Fe-rich clumpy structure at the edge of the remnant has been known about for some time  \citep[][]{1995ApJ...441..680V,2017ApJ...834..124Y}. Additionally the late-time optical spectra of SNe Ia show narrow absorption features that imply the possibility of large discrete clumps of high-velocity Fe-rich ejecta \citep{2019MNRAS.483.1114B}. Thus, clumps of Fe-rich ejecta may be a common feature of SNe Ia. To explain such a large fragment in SNe Ia, we may need some common mechanisms. 

\begin{figure}[t!]
 \begin{center}
  \includegraphics[bb=0 0 678 657, width=7cm]{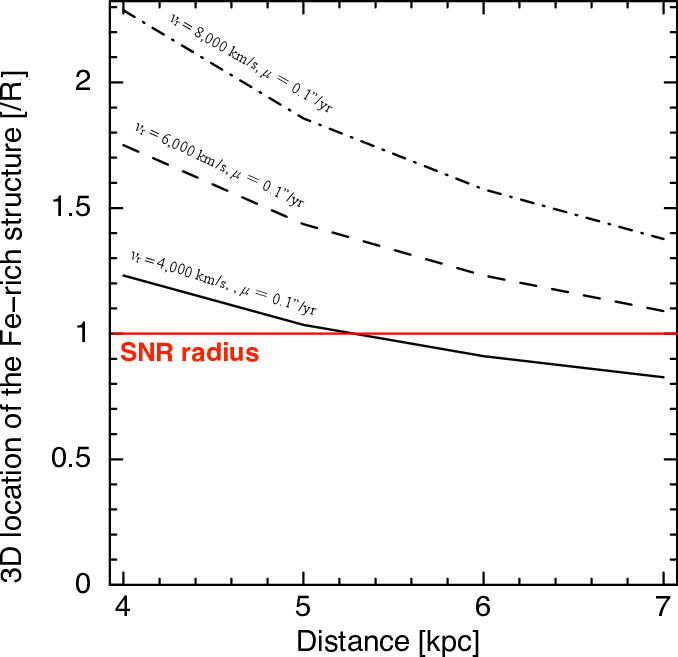}
 \end{center}
\caption{Estimation of 3D location of the Fe-rich structure in Kepler's SNR using eqn.~(\ref{eq:1}). The transverse velocity change with the distance to the remnant of 4--7 kpc. Here we assumed the proper motion of $\mu$ = 0.1$^{\prime\prime}$ yr$^{-1}$, the radial velocity of $v_{\rm r}$ = 4,000--8.000 km s$^{-1}$ \citep{2017ApJ...845..167S} and the projected radius of $r_{\rm pro}$ = 57$^{\prime\prime}$. The red line shows the typical SNR radius, $R$ = 1.8$^{\prime}$.}
\label{fig:3Dpos}
\end{figure}

Figure \ref{fig:KeplerTycho} shows a comparison of Fe-rich structures between Kepler's and Tycho's SNRs. In the case of Kepler, the Fe-rich structure is projected about halfway between the center and edge of the remnant in the image, although because of its high line-of-sight velocity the Fe-rich structure may be located close the remnant's outer blast wave in 3D. Proper motions at the northern rim of Kepler's SNR show values of  $\sim$0.8--0.11$^{\prime\prime}$ yr$^{-1}$ \citep{2008ApJ...689..225K,2008ApJ...689..231V} corresponding to speeds of  2,300--3,100 km s$^{-1}$ at the distance of 6 kpc \citep[e.g.,][]{2019arXiv190504475M}. The 3D velocities of associated knots SW1 and SW2 from \cite{2017ApJ...845..167S} are $\sim$5,000--8,000 km s$^{-1}$, significantly higher than the expansion velocity of the northern rim.  We can estimate the 3D position of the Fe-rich structure from the explosion center, $R_{\rm 3D}$ as
\begin{equation}\label{eq:1}
   R_{\rm 3D} = r_{\rm pro}/ \cos[\arctan(v_{\rm r}/v_{\rm tr})]
\end{equation}
where $r_{\rm pro}$, $v_{\rm r}$ and $v_{\rm tr}$ are the projected radius on the sky and the radial and transverse velocities of the Fe-rich structure. The transverse velocity depends on the proper motion, $\dot \theta$, and the distance, $D$, as $v_{\rm tr} = \dot \theta D$. 
In Figure \ref{fig:3Dpos}, we use these relations to estimate the 3D location of the structure based on previous measurements of the radial velocity and proper motion \citep{2017ApJ...845..167S}. The location of the Fe-rich structure lies outside the SNR radius in nearly all cases. These points support our contention that the Fe-rich structure in Kepler's SNR is located near the outer edge of the remnant similar to the Fe-rich structure in Tycho's SNR. The Fe-rich structure in Kepler's SNR is slightly larger (the solid angle is $\sim$0.1 sr) than the one in Tycho's SNR ($\sim$0.04 sr). On the other hand, Tycho's Fe-rich knot is surrounded by other Si-rich structures, which makes an even larger protruding structure \citep[Figure \ref{fig:KeplerTycho}, and see also][]{1995ApJ...441..680V,2017ApJ...840..112S,2017ApJ...834..124Y}. 

The biggest difference between these Fe-rich structures is in their element composition. Interestingly, \cite{2017ApJ...834..124Y} found no emission from Fe-peak elements other than Fe from this structure and concluded that either incomplete Si burning or the $\alpha$-rich freeze-out zone (with a peak temperature of $\sim$5.3--5.7 GK) would be the originating regime for the Fe-rich structure in Tycho's SNR. Therefore, considering the current work, it would appear that the Fe-rich structure in Tycho's SNR was produced at a somewhat deeper burning layer than the one in Kepler's SNR. Currently we do not know what can account for this difference. 



There are only a few theoretical studies predicting clumpy Fe-rich ejecta in SN Ia models. \cite{2005A&A...430..585G} simulated pure deflagration models with multiple ignition points, which provided four to five large $^{56}$Ni(=$^{56}$Fe) clumps at the time of maximum brightness \citep[see also][]{2000astro.ph..8463K,2002A&A...391.1167R,2003Sci...299...77G}. At a minimum, such Fe clumps appear to be a common feature in the deflagration phase. Here we suggest that such an effect by the deflagration flame in the initial stage might make the Fe-rich structures in Type Ia SNRs. 
However, in the case of the 3D pure deflagration models, the predicted sizes of Fe clumps are too large to explain the clumps in both Kepler and Tycho \citep[see Fig. 25 in][]{2005A&A...430..585G}.  In addition, ejecta clumps have been recognized as small-size structures also in SN-Ia observations. \cite{{2002ApJ...567.1037T}} investigated the maximum scale of clumping from the variation of absorption features in SN Ia spectra. They suggested $\sim$1\% perturbations of the photodisk area (the projection of the photosphere on the sky) would be consistent with the homogeneity of the Si II $\lambda$6355 absorption features.  This is  much smaller than the size of Fe clumps in the 3D pure deflagration models. 

Smaller scale Fe-rich clumps in SNe Ia have been realized in some delayed detonation models. \cite{2017ApJ...834..124Y} discussed asymmetric Fe distribution using the N100 model (a 3D delayed detonation model) of \cite{2013MNRAS.429.1156S}. The model produced some Fe clumps at the outer layer of the SNe Ia during the  explosion. Since large scale clumps break up when hit by the detonation wave, the large structures seen in the pure deflagration models do not appear here. The Fe clumps made in these models are almost freely expanding and can survive into the young SNR stage \citep[e.g.,][]{2019ApJ...877..136F,2019ApJ...879...64S}.
We do not expect the size difference of the Fe-rich structures between Kepler and Tycho (see Figure \ref{fig:KeplerTycho}) to be a significant issue in those models because such differences could result from slight differences in other conditions during the explosion. On the other hand, the Fe-rich structures produced in such an asymmetric deflagration are expected to have a large amounts of material from the n-NSE regime \citep[e.g.,][]{2000astro.ph..8463K,2013MNRAS.429.1156S,2017ApJ...834..124Y}, which is a significant different from the observational results for both Kepler and Tycho. 


Qualitatively, a detonation (supersonic wave) during the explosion would have a difficult time producing the sort of Fe-rich clumps observed in Type Ia SNRs. This is because the detonation wave passes through the WD so quickly that there would be little time for clumpy structures to grow. \cite{2010A&A...514A..53F} investigated the explosion of sub-$M_{\rm Ch}$ WDs via the double detonation scenario in two dimensions \citep[see also the recent 3D double-detonation model;][]{2018ApJ...868...90T}. The Fe ejecta are quite symmetrically distributed in these models, confirming our simple arguments on the difficulty of producing Fe-rich clumps from detonations. 


The asymmetric ejecta distribution expected in violent merger models may offer an explanation for the Fe-rich structure in the remnants \citep{2011A&A...528A.117P,2012ApJ...747L..10P}, however these models also produce large asymmetry in the distribution of the other elements. Such large asymmetry would be difficult to represent the symmetric layered materials as seen in Tycho's SNR \citep[e.g.][]{2010ApJ...725..894H}. In addition, the existence of the CSM produced by the companion wind \citep[e.g.,][]{2015ApJ...808...49K} could not support the WD merger senerio for the Kepler's supernova.
 Therefore, the merging models are difficult to reconcile with the observational properties of the SN Ia remnants.

Based on the preceding discussion, we suggest that the deflagration wave rather than detonation or WD mergers
offers a more plausible explanation for the origin of Fe-rich structures in Type Ia SNe and SNRs. On the other hand, there are, at present, no models that can explain all the characteristics of these Fe-rich structures, in particular their element composition. We hope future theoretical studies in Type Ia supernovae will shed more light on this problem.

\subsection{Nucleosynthesis for Other Type Ia SNRs} \label{otherSNR}
In this study of Kepler's SNR we have shown the value of focusing on local Fe-rich structures to interrogate the nuclear burning process in Type Ia SNRs. In section \ref{sec:nucleosynthesis}, we showed that the Fe-peak elements (and also Ca) are quite informative for understanding where distinct structures form deep in the interior of thermonuclear explosions.  Application to other remnants of this type is clearly desirable. Furthermore in the next decade and beyond we expect great progress through instrumental improvements in spectral resolution (thanks to X-ray calorimeters) and greatly enhanced telescope collecting area with missions such as {\it XRISM} \citep{2018SPIE10699E..22T}, {\it Athena} \citep{2018SPIE10699E..1GB}, and {\it Lynx} \citep{2019arXiv190609974F}. Higher sensitivity to detect faint lines will allow examination of the nuclear burning process in remnants of Type Ia (as well as core-collapse) supernovae in much greater detail. In the rest of this section, we highlght for a few other  Type Ia SNRs some topics related to Fe-group nucleosynthesis, guided by our nucleosynthesis models and with an eye toward future studies.

\vspace{0.2cm}
{\it 3C 397 \& Titanium, Chromium:} \cite{2015ApJ...801L..31Y} showed high Ni/Fe and Mn/Fe mass ratios in 3C 397, which indicates high neutronization in the supernova ejecta that can only be achieved by electron capture in the dense cores of exploding WDs (i.e., n-NSE) with a near-Chandrasekhar mass. 3C 397 is a bright, spatially extended X-ray remnant (covering a size of $\sim$5$^{\prime}$ $\times$ 3$^{\prime}$). Thus, it is possible to investigate the spatial distribution of Fe-peak elements in this remnant. Specifically, finding a structure with a low Ca/Fe mass ratio and and a high Ni/Fe ratio would be strong evidence for the presence of the n-NSE burning regime in the remnant. This can be seen in the n-NSE parts of Figure \ref{fig:SNeIamodel} (left panels for the near-$M_{\rm Ch}$) which correspond to peak temperatures $\gtrsim$ $8.5\times 10^9$ K (red and purple colored points). \cite{2005ApJ...618..321S} have presented a spatially resolved spectroscopic study of 3C 397 using {\it Chandra} observations \citep[see also][]{2010SCPMA..53S.267J}. They showed a low Ca/Fe abundance in both the Eastern ([Ca/Fe]/[Ca/Fe]$_\odot \sim 0.12$) and Western ([Ca/Fe]/[Ca/Fe]$_\odot \sim 0.21$)
Lobes of the remnant, which correspond to mass ratios of $\sim$0.4--0.7\%\footnote{\cite{2005ApJ...618..321S} did not quote a reference for the solar abundance values they used. We assume they used the \cite{1989GeCoA..53..197A} abundance values as in their earlier paper \citep{2000ApJ...545..922S}}. These low mass ratios imply an origin deeper into the incomplete Si burning than for the Kepler Fe-rich structure (i.e., slightly below horizontal band in the middle panels of Figure \ref{fig:SNeIamodel}). They also found spatial variation of the element distributions in the remnant. Measuring the Ni/Fe mass ratio at places where Ca/Fe is low and following the approach we have developed here, could provide evidence for the presence of material from the n-NSE regime. A more detailed spatially resolved spectroscopic study will be important for understanding nucleosynthesis in the explosion of 3C 397.

Future observations of Ti and Cr in 3C 397 will be interesting because the production of these elements in the n-NSE layer is very sensitive to the central density of the WD \citep[e.g.,][]{1997ApJ...476..801W,2017ApJ...841...58D,2018ApJ...861..143L}.
$^{50}$Ti and $^{52,54}$Cr are produced at the deepest layers in near-$M_{\rm Ch}$ explosions (see dashed green and solid purple curves in Figure \ref{fig:nucleosynthesis}). The large amount of $^{50}$Ti and $^{52,54}$Cr at the core thanks to electron capture offers the possibility of obtaining additional strong evidence for a near-$M_{\rm Ch}$ explosion. At higher central densities, the production of these elements increases. In \cite{2015ApJ...801L..31Y}, their models with central density of 3 $\times$10$^{9}$ g cm$^{-3}$ required relatively high metallicities Z $\sim$ 5.4 Z$_\odot$ in order to reproduce high mass ratios for both Ni/Fe and Mn/Fe. On the other hand, if the central density were sufficiently high (5--6 $\times$10$^{9}$ g cm$^{-3}$), then the high Mn/Fe and Ni/Fe mass ratios in 3C 397 could be reproduced even with a normal progenitor metallicity (1--1.5 Z$_\odot$) \citep{2017ApJ...841...58D,2018ApJ...861..143L}. A test of whether the progenitor WD of the 3C 397 explosion had such a high central density would be finding a Ti and Cr rich region in the SNR.  Such a finding would allow an estimate for the central density of the progenitor to be made.

\vspace{0.2cm}
{\it W49B \& Nickel:} Recently, \cite{2018A&A...615A.150Z} discussed a Type-Ia supernova origin for W49B. They found a high Mn/Cr mass ratio of $\sim$ 1.3 (0.8--2.2) and estimated a rather high metallicity of Z = 0.12$^{+0.14}_{-0.07}$ = 8$^{+10}_{-4}$ Z$_{\odot}$ for the progenitor, assuming the incomplete Si burning regime in a Type Ia explosion. They also suggested the possibility of a combination of both incomplete Si burning and neutron-rich NSE (``normal" freeze-out) material in order to explain their overabundance of manganese. We note that these authors did not discuss Ni, which is arguably more important for understanding the burning regime and the progenitor metallicity.

From the {\it XMM-Newton} spectrum of the bright central region in W49B,
\cite{2006A&A...453..567M} derived a large overabundance of nickel, Ni/Ni$_{\odot} = 10^{+2}_{-1}$ (using the solar values in \cite{1989GeCoA..53..197A}).  Using their results we estimate that the observed Ni/Fe mass ratio is not so high, $\lesssim$ 8\% (90\% upper limit).   W49B is now known to be in an overionized state through {\it Suzaku} observations \citep{2009ApJ...706L..71O}, which tends to reduce the fitted Ni abundance from the values derived with the spectral models used by \cite{2006A&A...453..567M}.
%
For an overionized plasma, there is a relatively strong Fe-He$\beta$ cascade line (principle quantum number n = 3 $\rightarrow$ 1) close to the energy of the Ni-K$\alpha$ line. In spectral models that do not include the overionization condition, the flux of the Fe-He$\beta$ cascade line is incorrectly attributed to Ni-K$\alpha$.  When \cite{2009ApJ...706L..71O} tested the use of a collisional ionization equilibrium (CIE) spectral fit, they too derived a large Ni/Ni$_{\odot}$ abundance ratio of $\sim$10.9 from their {\it Suzaku} data, which is consistent with that measured by {\it XMM-Newton}. On the other hand, by considering the recombination lines and continuum, they estimated a smaller abundance of nickel, Ni/Ni$_{\odot} \sim$ 5.2 (almost half the CIE result). Thus, the actual Ni/Fe mass ratio in the bright central region should also be smaller, $\lesssim$ 4\% (again, assuming half the CIE result). The {\it NuSTAR} observations also detected the Ni emission at both  the east and west sides of the remnant \citep{2018ApJ...868L..35Y} where the observed Ni/Fe mass ratio is estimated to be $\sim$4--5\%. 

\begin{figure}[h!]
 \begin{center}
  \includegraphics[bb=0 0 784 780, width=8cm]{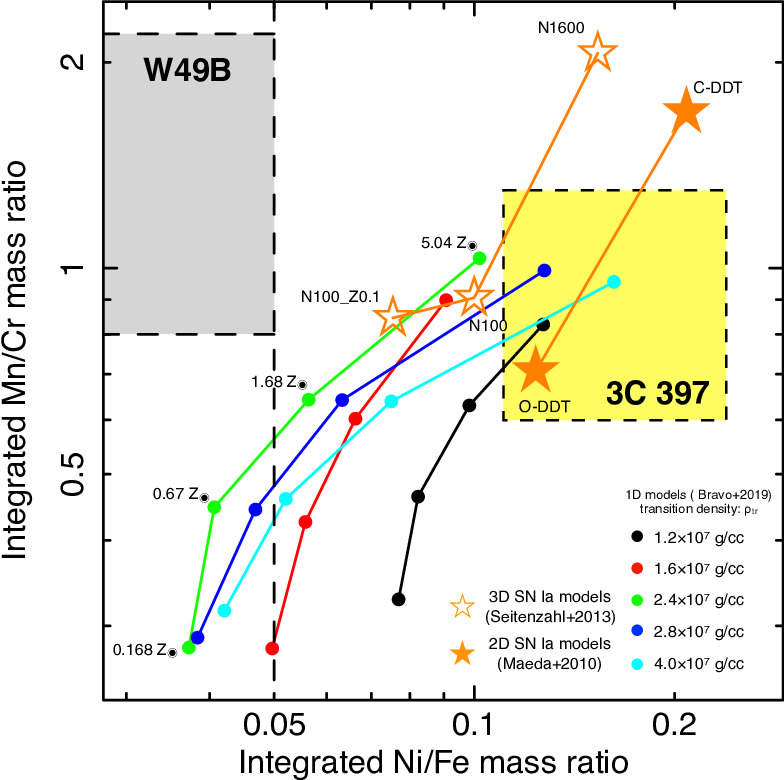}
 \end{center}
\caption{A scatter plot between the Mn/Cr and Ni/Fe mass ratios. The data plots of mass ratios were estimated from the integrated SN ejecta in Type Ia models. The filled gray area and yellow area show the observed values for W49B and 3C 397, respectively. The filled circles show the 1D models in \cite{2019MNRAS.482.4346B}. The filled and hollow stars show the 2D \citep{2010ApJ...712..624M} and 3D \citep{2013MNRAS.429.1156S} models, respectively.}
\label{fig:w49b_3c397}
\end{figure}

Here we emphasize the relatively small amount of Ni with respect to Fe (the observed Ni/Fe mass ratio is $\sim$ 5\% at most) in the remnant, which is difficult to explain with our Type Ia nucleosynthesis models. Although the value of the mass ratios of Mn/Cr$\sim$1 and Ni/Fe $<$ 0.05 naively suggest an origin in the incomplete Si burning zone and in particular to the lowest peak temperatures shown in the top and bottom panels of Figure \ref{fig:SNeIamodel}, in fact such a scenario would require that the reverse shock in W49B to have propagated only to a mass coordinate of $\sim$1 $M_\odot$ so that only the outermost portion of the ejecta was emitting X-rays. This is inconsistent with estimates of the age of W49B \citep[5--6 kyr:][]{2018A&A...615A.150Z} and  the presence of very bright Fe-K emission, which requires that most of the ejecta be shock-heated already \citep[e.g.,][]{2015ApJ...803..101P}.  In the case of a progenitor with super-solar metallicity, such as Z = 8$^{+10}_{-4}$ Z$_{\odot}$, the Ni/Fe mass ratio in the NSE and n-NSE should be much higher than a few percent. 
%
This can be seen in Figure \ref{fig:SNeIamodel} (bottom left), where the high-metallicity model plotted (with Z = 5.04 Z$_{\odot}$) shows Ni/Fe mass ratios close to $\sim$20\% in the NSE region. 
Therefore, the observed small amount of observed Ni does not support such an extremely high metallicity case. In addition, if the n-NSE layer in W49B is already heated, there should be the Fe-rich region that is colocated with a large amount of Ni with Ni/Fe mass of $\gtrsim$ 10\% (see $>$ 7 GK region in the bottom left plot in Figure \ref{fig:SNeIamodel}). However, there is no trace of such a structure in the remnant. Also in this case, the total amount of Ni in the whole remnant should be as large as 3C 397 \citep[Ni/Fe = 11--24\%;][]{2015ApJ...801L..31Y}, which is factors of a few larger than the mass limits in W49B. 

Next we consider the mass ratios estimated from the spectrum of the whole remnant and compare to the yields integrated over the entire ejecta. In Figure \ref{fig:w49b_3c397}, we plot the Mn/Cr and Ni/Fe mass ratios for W49B (gray box) and 3C 397 (yellow box). The Mn/Cr ratios are comparable to each other, however the Ni/Fe mass ratio in W49B is significantly less than that in 3C 397: $\lesssim$ 5\% vs. 11--24\%. We also plot mass ratios from some SN Ia models (various curves plotted with different colors). This figure shows that no existing  Type Ia model can reproduce the low integrated Ni/Fe mass ratio in W49B. In particular, the models that can explain the observed element abundance of W49B in \cite{2018A&A...615A.150Z} (e.g., N100, N1600, O-DDT) have much larger Ni/Fe mass ratios. Therefore, the high Mn/Cr mass ratio can not be necessarily caused by the n-NSE layer in the near-$M_{\rm Ch}$ explosions. Thus, we argue that accurate measurements of the Ni abundance in W49B are necessary to arrive at a definitive conclusion on its origin.

We note that a SN Ia origin for this object would also require an explanation for its large Fe K$\alpha$ centroid and luminosity, which are among the highest in the sample of X-ray bright SNRs \citep{2014ApJ...785L..27Y,2018ApJ...865..151M}. To produce such a large ionization timescale at the estimated age and radius of W49B would require a very strong CSM interaction \citep{2007ApJ...662..472B}, which would make this object unique among Type Ia SNRs.

In the case of core-collapse models, there can be a large variation in the yields depending on the electron fraction $Y_{\rm e}$ at the burning layer. In many studies, the $Y_{\rm e}$ value is fixed in the calculation to model a specific observational abundance pattern and the actual $Y_{\rm e}$ value may not be indicated. For example, the core-collapse models of \cite{2006NuPhA.777..424N} that are referenced in \cite{2018A&A...615A.150Z} assumed a high electron fraction $Y_{\rm e}$ = 0.4997 at the incomplete Si burning zone in order to explain the abundance patterns of metal-poor stars \citep[see also][]{2005ApJ...619..427U}, which produced a low abundance of Mn in the models.  Considering the uncertainties in  $Y_{\rm e}$ might change the impression of Fig. 9 in \cite{2018A&A...615A.150Z}.

\begin{figure}[t!]
 \begin{center}
  \includegraphics[bb=0 0 997 758, width=8cm]{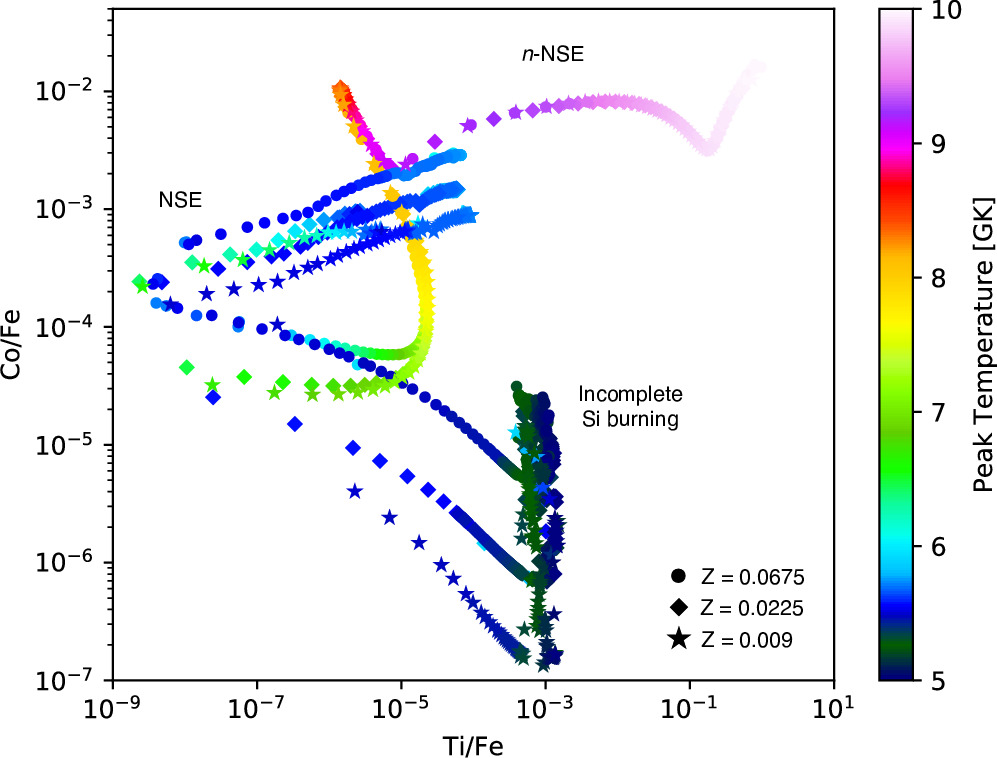}
 \end{center}
\caption{The model prediction of the Ti/Fe amd Co/Fe mass ratios. The model ssumed the delayed detonation with the transition density, $\rho = 2.8 \times 10^{7}$ g cm$^{-3}$ and the metallicity of Z = 0.009 (filled star), Z = 0.0225 (filled diamond) and Z = 0.0675 (filled circle).}
\label{fig:TiCo}
\end{figure}

\vspace{0.2cm}
{\it Tycho's SNR \& Titanium:} 
\cite{2015ApJ...805..120M} claimed detection of a shocked Ti line at an energy of $\sim$4.9 keV in Tycho's SNR using {\it XMM-Newton} data. Their results indicated that the shocked Ti was spatially colocated with other iron-peak nuclei.
However, \cite{2017ApJ...834..124Y} reported no detection of this line with {\it Suzaku}. In addition, the centroid energy of the purported Ti line indicated a hydrogen-like charge state rather than the more plausible helium-like or lower charge state.   Also, the precise values of the fit parameters for the thermal plasma spectral model (i.e., ionization age and electron temperature) can influence the intrinsic emissivity of the Ca He$\gamma$ line at $\sim$4.9 keV, which may be the origin of the line structure. 
The typical Ti/Fe mass ratio at the incomplete Si burning regime is $\sim$0.1\% (see Figure \ref{fig:TiCo}), so verifying the existence of shocked Ti line emission will likely need to wait for much deeper X-ray spectra.

Detecting radioactive $^{44}$Ti would also be interesting for Tycho's SNR.
Actually, $^{44}$Ti is not produced sufficiently in either of our near-$M_{\rm Ch}$ delayed detonation or sub-$M_{\rm Ch}$ detonation models, whereas double-detonation models with a massive He shell of $\sim$0.1--0.2 $M_{\odot}$ predict a large amount of $^{44}$Ti \citep[e.g.,][]{1994ApJ...423..371W,1995ApJ...452...62L,2010A&A...514A..53F}. For these double-detonation models, the $^{44}$Ti is produced by the He detonation, and those elements (also including a large amount of Fe) should be distributed around the outer edge of the ejecta.
Thus, the spatial distribution of $^{44}$Ti will be quite different from that in core-collapse supernovae \citep[e.g.,][]{2017ApJ...834...19G}.

The detection of $^{44}$Ti in Tycho's SNR is still controversial.
\cite{2014ApJ...789..123W} reported a bump in the 60--90 keV energy band by {\it INTEGRAL}, potentially associated with the $^{44}$Ti line. There is also the potential detection of the same hard X-ray features by the {\it SWIFT}/BAT \citep{2014ApJ...797L...6T}. On the other hand, {\it NuSTAR} observations showed no evidence for these $^{44}$Ti lines and set an upper-limit on the mass of $^{44}$Ti of $<$ 2.4 $\times$ 10$^{-4} M_{\odot}$ for a distance of 2.3 kpc \citep{2015ApJ...814..132L}. In addition, there is no trace of the He-detonation shell in this remnant, which would support this small amount of synthesized $^{44}$Ti. Future observations by X-ray calorimeters may be able to set a tight upper limit on the $^{44}$Ti mass using the cascade line from $^{44}$Sc ($\sim$4.1 keV for neutral, $\sim$4.3 keV for He like). In addition to Tycho's SNR, searching for the $^{44}$Sc cascade line in the young SNR G1.9+0.3 \citep{2010ApJ...724L.161B} should be fruitful with X-ray calorimeter missions as long as Doppler velocity smearing of the line is not too extreme.

\vspace{0.2cm}
{\it Cobalt in Type Ia Ejecta:} Co will also be an interesting species for tests of Type Ia nucleosynthesis at deeper layers. $^{59}$Co is produced as $^{59}$Cu and $^{59}$Ni at the $\alpha$-rich freeze-out and as $^{59}$Co and $^{59}$Ni in neutron-rich NSE layers (dashed blue curve in Figure 1 and see also Iwamoto et al. 1999). This is a neutron-rich element in origin (as compared to iron, mainly coming from $^{56}$Ni), thus the amount synthesized depends on the extent of neutronization from both the progenitor's metallicity and electron capture. Thus, this element is produced similarly to $^{58}$Ni. On the other hand, the K-shell emission from Co will appear in a complex region around 6.9--7.2 keV where the Fe Ly$\alpha$ and K$\beta$ emissions exist. Therefore, X-ray calorimeter missions with large effective area (e.g., {\it Athena}, {\it Lynx}) will offer the best chance for detection. 

Figure \ref{fig:TiCo} summarizes the model prediction for the Ti/Fe and Co/Fe mass ratios. From regime to regime, the production of each element differs from each other. Here we can see clearly that the Co/Fe mass ratio has the stronger metallicity dependence. The Co/Fe mass ratio at the NSE regime is $\lesssim$0.3\%. At the core of exploding WDs, the Co/Fe mass ratio grows to $\sim$0.2--2\%.

\section{Summary and Conclusion} \label{sec:future}
Iron in SNe Ia is mainly produced at inner layers (incomplete Si burning, NSE, n-NSE) of the explosion, which means that the Fe-rich structures in Type Ia SNe and SNRs carry information on the interior properties of exploding WDs. The mass fractions among the Fe-peak elements at these layers vary from layer to layer.
Therefore, by measuring the mass fraction in observed Fe-rich structures, we can determine the origin of these features and probe nucleosynthesis at interior burning layers of the explosion. We have demonstrated this capability here for an Fe-rich structure in  Kepler's supernova remnant.

{\it Chandra} X-ray observations reveal the existence of a high-speed Fe-rich ejecta structure moving a $\sim$5,000--8,000 km s$^{-1}$ at the southwestern region in the remnant. Such high velocities that exceed the mean expansion velocity of the rim means the Fe-rich structure is located close to the outermost extent of the remnant. We found strong K-shell emissions from Fe-peak elements (Cr, Mn, Fe, Ni) and calcium in the Fe-rich structure for the first time. From these, we determined Ca/Fe, Cr/Fe, Mn/Fe and Ni/Fe mass ratios of 1.0--4.1\%, 1.0--4.6\%, 1--11\% and 2--30\%, respectively. All of these observational mass ratios are consistent with a narrow range of peak temperature ($\sim$(5.0--5.3)$\times$10$^{9}$ K) within the incomplete Si burning layer for a progenitor with super-solar metallicity, Z $>$ 0.0225. Thus, we conclude that most of the ejecta in the Fe-rich structure was processed by incomplete-Si-burning.

At present, we do not have a clear understanding of how to produce such a distinct structure of incomplete-Si-burning ejecta close to the edge of remnant. Theoretical studies suggest some mechanisms for producing Fe clumps in the initial stage of the explosion (most likely during the deflagration phase), however there are no accurate predictions of the properties of such structures. We hope that future theoretical studies will reveal more about clumping in SN Ia explosions. 

We also discussed future prospects on Type Ia nucleosynthesis with planned and future X-ray calorimeter missions. The optimal measurements of faint X-ray lines (e.g., Ni, Ti and Co) from local spots in Type Ia SNRs will require high spectral-resolution, large collecting areas and arcsecond imaging.  Such advanced spectra have the potential to reveal much about SN Ia explosions and their progenitors.

\acknowledgments{T.S.\ was supported by the Japan Society for the Promotion of Science (JSPS) KAKENHI Grant Number JP18H05865, JP19K14739, and the Special Postdoctoral Researchers Program and FY 2019 Incentive Research Projects in RIKEN. J.P.H acknowledges support for X-ray studies of SNRs from NASA grant NNX15AK71G to Rutgers University. E.B. acknowledges support from the MINECO-FEDER grants AYA2015-63588-P and PGC2018-095317-B-C21. We also thank the anonymous referee for comments that helped us to improve the manuscript.
}

\software{CIAO \citep[v4.10;][]{2006SPIE.6270E..1VF}, Xspec\citep[v12.10.0c;]{1996ASPC..101...17A}}

\appendix
\section{Detailed Analysis of Fe-rich Structure}
The Fe-rich structure we focused on in this study is located at a little inside in the 2D image, probably due to projection effects. Therefore, the element composition in the region may have contamination from the area overlapping on the line of sight. Our conclusion (the incomplete-Si-burning origin for the Fe-rich structure) crucially depends on the observed Ca/Fe (Figure \ref{fig:restricted_T}). If the observed element abundances, in particular Ca/Fe, were highly contaminated from the other structures, it would be difficult to support this conclusion. Thus, we here show a more detailed analysis for the Fe-rich structure to strengthen our interpretation.

\begin{figure}[h!]
 \begin{center}
  \includegraphics[bb=0 0 964 481, width=16cm]{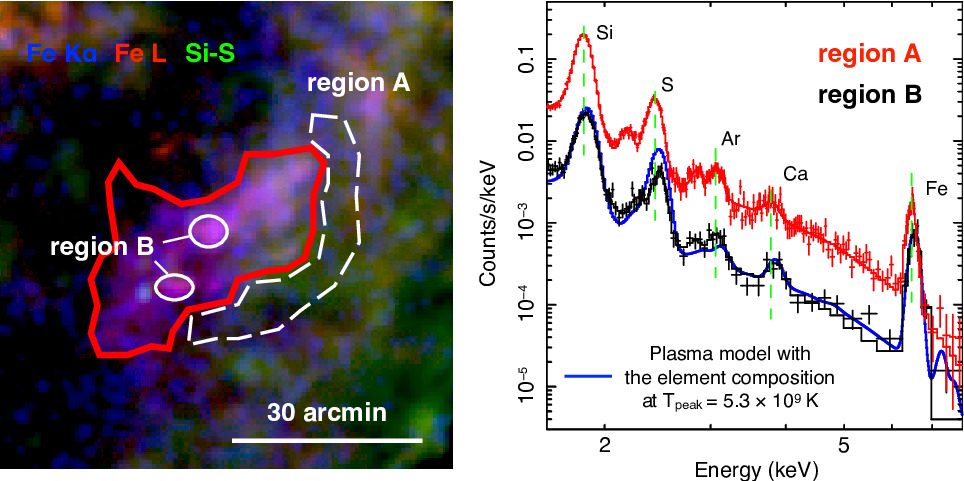}
 \end{center}
\caption{Left: the enlarged image around the Fe-rich structure at the southwestern region. Right: X-ray spectra in region A (red) and B (black). The blue curve shows a spectral model (vvpshock + power law) with the element composition at the incomplete Si burning layer ($T_{\rm peak} = 5.3 \times 10^{9}$ K). The mass ratios of Si/Fe, S/Fe, Ar/Fe and Ca/Fe at this layer are 0.013, 0.019, 0.009, and 0.017, respectively. We assumed a 6 keV plasma model with a blue-shifted velocity of $\sim$5,000 km s$^{-1}$. The black and red lines show best-fit models consisting of a power law + several Gaussian models.}
\label{fig:ap}
\end{figure}

We extracted spectra from region A and B (Figure \ref{fig:ap}). Region A is adjacent to the Fe-rich region we defined in Figure~\ref{fig:f1}, where the emission should be similar to the contamination from the overlying outer layers. Region B consists of the brightest knots in the Fe-rich structure where the spectrum with less contamination should show the elemental composition of this structure more purely. The comparison between them will help us to understand the contamination in the Fe-rich structure.

First, we found the X-ray lines in these regions have different centroid energies. In particular, we can see clearly that both Si K$\alpha$ and Fe K$\alpha$ have higher centroid energies than those in region A, which implies the lighter elements are also moving together with the iron. In fact, the line centroids are well fitted with a blue-shifted velocity of $\sim$5,000 km s$^{-1}$ (Figure \ref{fig:ap} right). If these lighter elements are pure contamination from the other regions, the line centroids would be different as in region A.
Therefore, the line centroids would support the same origin between the iron and the other lighter elements.

Second, we found the element abundances including lighter elements in region B are very similar to those in the incomplete Si burning layer of our SN Ia models. The blue curve in the right panel of Figure~\ref{fig:ap} shows a spectral model assuming the elemental composition at the incomplete Si burning layer. Here, in particular, both silicon and calcium are fitted very well by the model composition. At the n-NSE layer, we need almost zero abundance for both silicon and calcium (see Figure \ref{fig:nucleosynthesis}), which implies that the n-NSE origin for this structure would not be reasonable. Thus, the elemental abundances including the lighter elements  can also support the incomplete-Si-burning origin for the Fe-rich structure.

Based on these results, we conclude the contamination from the other structures to the Fe-rich structure is not significant and does not change our interpretation.

\end{document}